\documentclass[pra,a4paper,twocolumn,floatfix]{revtex4}

\usepackage{graphicx,amssymb}

\begin{document}

\title{Ultrafast electron dynamics following outer-valence ionization: The impact of low-lying relaxation satellite states}

\author{Siegfried \surname{L\"unnemann}}
\email[Electronic mail: ]{siegfried.luennemann@pci.uni-heidelberg.de}
\author{Alexander I. \surname{Kuleff}}
\email[Electronic mail: ]{alexander.kuleff@pci.uni-heidelberg.de}
\altaffiliation[On leave from: ]{Institute for Nuclear Research and Nuclear Energy, BAS,
                                72, Tzarigradsko Chaussee Blvd., 1784 Sofia, Bulgaria}
\author{Lorenz S. \surname{Cederbaum}}
\affiliation{Theoretische Chemie, PCI, Universit\"at Heidelberg\\
            Im Neuenheimer Feld 229, 69120 Heidelberg, Germany}

\date{\today}

\begin{abstract}
Low-lying relaxation satellites give rise to ultrafast electron dynamics following outer-valence ionization of a molecular system is studied. To demonstrate the impact of such satellites, the evolution of the electronic cloud after sudden removal of an electron from the highest occupied molecular orbital (HOMO) of the organic unsaturated nitroso compound 2-Nitroso[1,3]oxazolo[5,4-d][1,3]oxazole is traced in real time and space using \textit{ab initio} methods only. Our results show that the initially created hole charge remains stationary but on top of it the system reacts by an ultrafast $\pi-\pi^*$ excitation followed by a cyclic excitation-deexcitation process which leads to a redistribution of the charge. The $\pi-\pi^*$ excitation following the removal of the HOMO electron takes place on a sub-femtosecond time scale and the period of the excitation-deexcitation alternations is about 1.4~fs. In real space the processes of excitation and de-excitation represent ultrafast delocalization and localization of the charge. The results are analyzed by a simple two- and three-state model.
\end{abstract}

\maketitle

\section{Introduction}

With the advent of the attosecond pulse techniques (see, e.g. Refs. \cite{CorKra07,Reider04} and the references therein) the scientific community obtained a powerful tool to monitor the electron dynamics in atomic and molecular systems and to study processes that take place on a time scale in which the electronic motion is still disentangled from the slower nuclear dynamics. Such kind of processes are, for example, the response of the electronic cloud to an ultrafast perturbation, like ultrafast excitation or ionization \cite{Drescher,Nest07,Nest08,Bandrauk08,Scrinzi}. About ten years ago in a first work \cite{Zobeley99} it was shown that indeed after a sudden ionization rich ultrafast electron dynamics may occur. The positive charge created after the ionization can migrate throughout the system solely driven by the electronic many-body effects -- the electron correlation and electron relaxation \cite{Zobeley99,Breidbach03}. Since this charge migration is ultrafast, typically few femtoseconds \cite{Breidbach03,Hennig05,Kuleff05,Kuleff07,Breidbach07}, it can be calculated neglecting the nuclear motion as long as one is concerned with the relevant time interval during which this ultrafast process takes place. Clearly, at later times the nuclear dynamics will come into play and will perturb the picture. Thus, if we wish to know precisely what happens at this later stage, the nuclear motion must be considered. However, since several or even many electronic states participate, an adequate description of the nuclear motion is rather involved. It should be noted, in this respect, that very recently a time-dependent Born-Oppenheimer approximation for treating quantum mechanically coupled electron-nuclear dynamics was proposed \cite{Cederbaum08,Miller08} which might provide a possibility to attack such complicated problems.

We have studied the charge migration in different molecules \cite{Breidbach03,Hennig05,Kuleff05,Kuleff07,Breidbach07} and found that it usually takes place after ionization in the inner-valence shell where the electron correlation effects are typically much stronger than in the outer-valence shell. However, very recently we have shown that the charge migration phenomenon is not inherent only to the inner-valence ionized states but can take place also after an outer-valence ionization \cite{Luennemann08}. The system studied in Ref. \cite{Luennemann08} possesses a chromophore-donor site which is initially ionized and the created hole charge migrates throughout the system to the amine-acceptor site within just 4 fs. Further analysis \cite{Luennemann_JCP08} then showed that the mechanism underlying this ultrafast migration is the so-called hole mixing \cite{NiBiSciCe82} which is one of several ways of the manifestation of electronic many-body effects in the ionization process \cite{CedDoSchNie}.

In the present work we want to continue the investigation of ultrafast electron dynamics following ionization of the outer-valence shell focusing on a different mechanism, namely the so-called dominant satellite mechanism. It is known from the early days of the photoelectron spectroscopy that the correlation and relaxation effects can lead to the appearance of additional weak bands in the photoelectron spectrum, the so-called shake-up or satellite bands, which correspond to excitation processes accompanying the ionization \cite{Siegbahn}. This kind of shake-up states are typical for the inner-valence and core ionization and appear in the ionization spectra of nearly every atom or molecule. However, numerous theoretical and experimental studies showed that there are certain classes of compounds where satellites can appear also in the outer-valence region. These are systems containing heavier atoms, like transition-metal complexes \cite{NiesCed81,MonNies85}, or $\pi$-electron systems \cite{Masuda90,Wardermann92,WeikCed95,Deleuze,Ehara_MolPhys06}, which have low-energy virtual orbitals. Here we will concentrate on the latter class of systems and will investigate the impact of the low-lying satellites on the electron dynamics following outer-valence ionization of some organic unsaturated nitroso compounds \cite{Wardermann92}.

The paper is organized as follows. In Sec. \ref{sec_theory} the theoretical background of the methodology used for calculating the ultrafast electron dynamics is briefly outlined (Sec. \ref{basics}) and the different mechanisms of charge migration are briefly discussed (Sec. \ref{CM_mechanisms}). In Sec. \ref{sec_results} we present the results of our calculations including the ionization spectrum of the molecule 2-Nitroso[1,3]oxazolo[5,4-d][1,3]oxazole and the electron dynamics following ionization of its highest occupied molecular orbital (HOMO). Section \ref{sec_discussion} is devoted to the discussion and analysis of the results obtained via a simple two- and three-state model.

\section{Theoretical Background}\label{sec_theory}

\subsection{Basic equations}\label{basics}

In this section we briefly review the theoretical background of the methodology used to study ultrafast electron dynamics following ionization of a system. For technical details we refer the reader to Refs \cite{Breidbach03,Kuleff05,Breidbach07}.

The starting point of our investigation is a neutral molecule in its ground state $|\Psi_0\rangle$. The ionization of the system brings it into a non-stationary state $|\Phi_i\rangle$. A convenient quantity then for tracing the succeeding electron dynamics is the density of the so created initial hole which can be defined by the following expression:
\begin{equation}\label{eq1}
Q(\vec r,t) := \langle\Psi_0|\hat{\rho}(\vec r,t)|\Psi_0\rangle -
\langle\Phi_i|\hat{\rho}(\vec r,t)|\Phi_i\rangle =
\rho_0(\vec r) - \rho_i(\vec r,t),
\end{equation}
where $\hat{\rho}$ is the electron density operator. The first term on the right-hand side of Eq. (\ref{eq1}), $\rho_0$, is the ground state density of the neutral system and is time-independent. The second term, $\rho_i$, is the density of the cation and hence is time dependent, since $|\Phi_i\rangle$ is not an eigenstate of the cationic system. The quantity $Q(\vec r,t)$, referred hereafter as the \textit{hole density}, describes the density of the hole at position
$\vec r$ and time $t$ and by construction is normalized at all times $t$.

In the Heisenberg picture, the time-dependent part $\rho_i(\vec r,t)$ reads:
\begin{equation}\label{eq_rho_i}
\rho_i(\vec r,t) = \langle\Phi_i|e^{i\hat{H}t}\hat{\rho}(\vec r,0)
                   e^{-i\hat{H}t}|\Phi_i\rangle
= \langle\Phi_i(t)|\hat{\rho}(\vec r,0)|\Phi_i(t)\rangle,
\end{equation}
where $|\Phi_i(t)\rangle = e^{-i\hat{H}t}|\Phi_i\rangle$ is the propagating multielectron wavepacket.

Using the standard representation of the density operator in a one particle basis ${\varphi_p(\vec r)}$ and occupation numbers ${n_p}$, Eq. (\ref{eq1}) can be rewritten as follows
\begin{equation}\label{eq_Q_N}
Q(\vec r,t) = \sum_{p,q}\varphi_p^\ast(\vec r)\varphi_q(\vec r) N_{pq}(t).
\end{equation}
where the matrix $\textbf N(t) = \{N_{pq}(t)\}$ with elements
\begin{equation}\label{eq_N_t}
N_{pq}(t) = \delta_{pq}n_p - \sum_{M,N} \langle\Phi_i(t)|\tilde\Psi_M\rangle
            \rho_{MN} \langle\tilde\Psi_N|\Phi_i(t)\rangle
\end{equation}
is referred to as the hole density matrix. The second term of Eq. (\ref{eq_N_t}) is obtained by inserting in Eq. (\ref{eq_rho_i}) a resolution of identity of a complete set of appropriate ionic eigenstates $|\tilde\Psi_M\rangle$ before and after the density operator $\hat{\rho}(\vec r,0)$. The matrix $\rho_{MN}$ is thus the representation of the density operator within this basis.

Diagonalization of the matrix $\textbf N(t)$ for fixed time points $t$ yields the following expression for the hole density
\begin{equation}\label{eq_Q_n}
Q(\vec r,t) = \sum_p|\tilde\varphi_p(\vec r,t)|^2\tilde n_p(t),
\end{equation}
where $\tilde\varphi_p(\vec r,t)$ are called \textit{natural charge orbitals}, and $\tilde n_p(t)$ are their \textit{hole
occupation numbers}. The hole occupation number $\tilde n_p(t)$ contains the information on which part of the created hole charge is in the natural charge orbital $\tilde\varphi_p(\vec r,t)$ at time $t$. Because of the charge conservation, one finds that $\sum_p \tilde n_p(t) = 1$ at any time $t$.

For calculating the hole density and its constituents we use \textit{ab initio} methods only. The whole calculation consists of four steps. After determining the molecular geometry, the first step is a Hartree-Fock (HF) calculation. The second step is the calculation of the relevant part of the ionization spectrum via Green's function formalism. A computationally very successful approach to obtain the Green's function is the \textit{algebraic diagrammatic construction} [ADC($n$)] scheme \cite{Schirmer83}. In the present calculation we used the non-Dyson ADC(3) method \cite{Schirmer98,TrofSchir05} realized within the so-called intermediate-state representation \cite{Mertins96,Schirmer91}, an effective many-body basis serving as $|\tilde\Psi_M\rangle$ introduced in Eq. (\ref{eq_N_t}). The third step is the propagation of the multielectron wavepacket of the ionized system \cite{Kuleff05} with the help of the short iterative Lanczos technique \cite{Leforestier}. The fourth and last step is to build through Eq. (\ref{eq_N_t}) the matrix $\textbf{N}(t)$ and to diagonalize it in order to obtain the natural charge orbitals $\tilde{\varphi}_p(\vec r,t)$ and the hole occupation numbers $\tilde{n}_p(t)$, see Eq. (\ref{eq_Q_n}). With the help of these quantities we can now trace the evolution of the hole density of a system after suddenly removing one of its electrons.

At this point we would like to comment briefly on the choice of the initial state $|\Phi_i\rangle$. The above sketched methodology is independent of the particular choice and the way of preparation of the initial state as long as the ionized electron is removed from the system on a shorter time scale than that of the charge migration. The assumption made is that the initially created ionic state can be described by a separable manyelectron wavefunction, i.e. the interaction between the ionized electron and the remaining core is neglected -- sudden approximation (see, e.g. \cite{Pickup77}). Within this approximation the initial hole is described by the so-called Dyson orbital, i.e. the overlap between the $N$-electron initial and $(N-1)$-electron final wavefunctions. However, in the outer-valence region the Dyson orbitals differ very little from the canonical Hartree-Fock orbitals \cite{Nicholson99,Brion01}. That is why, in the numerical calculations to be discussed in this paper the initial state is prepared through a sudden removal of an electron from a particular HF-orbital. The initial state can, of course, be constructed such that it corresponds to a removal of an electron from a linear combination of HF-orbitals. This liberty allows one to reproduce the hole density of practically every particular initial vacancy. However, to avoid investigating many linear combinations of HF-orbitals of interest we concentrate on specific HF-orbitals. In this way we can unambiguously identify the basic mechanisms leading to charge migration. Since the time-dependent Schr\"odinger equation which governs the electron dynamics is a linear equation, these mechanisms are also operative when other choices of the initial state are used.

\subsection{Ionic states and mechanisms of charge migration}\label{CM_mechanisms}

Here we will briefly review the basic mechanisms of charge migration referring the interested reader to Ref. \cite{Breidbach03} for more details.

For proper understanding of the basic mechanisms of ultrafast charge migration following ionization it is illuminative to analyze a typical ionization spectrum. The calculated cationic spectrum consists of vertical lines, where each line represents a cationic eigenstate $|I\rangle$. The position of the line is given by the ionization energy, and its hight -- by the square of the transition amplitude $\langle \Phi_i | I \rangle$, a quantity related to the ionization cross section.  For the ease of interpretation we will expand the exact cationic state $|I\rangle$ in a series of electronic configurations, as is traditionally done in configuration interaction (CI) calculations (see, e.g. Ref. \cite{Szabo}):
\begin{equation}\label{cat_state}
|I\rangle = \sum_j c_j^{(I)} \hat a_j|\Psi_0\rangle + \sum_{a,k<l} c_{akl}^{(I)} \hat a_a^\dagger \hat a_k \hat a_l |\Psi_0 \rangle + \cdots ,
\end{equation}
where $|\Psi_0\rangle$ is the exact ground state of the neutral system, $\hat a$ and $\hat a^\dagger$ are the annihilation and creation operators, respectively, and $c^{(I)}$'s are the expansion coefficients. The indices $a,b, \ldots$ refer to unoccupied (virtual) orbitals (or particles), whereas the indices $i,j,\ldots$ indicate occupied orbitals (or holes). Throughout the whole text $p,q, \ldots$ will be used as general indices. Accordingly, the terms $\hat a_j|\Psi_0\rangle$ are called one-hole (1h) configurations, since one electron has been removed from the corresponding occupied orbital, the terms $\hat a_a^\dagger \hat a_k \hat a_l |\Psi_0 \rangle$ are referred to as two-hole-one-particle (2h1p) configurations, indicating that in addition to the removal of one electron another one is excited to a virtual orbital, and so forth. Note that in the spirit of the GF approach, which accounts also for the ground-state correlations, in Eq. (\ref{cat_state}) the expansion is applied on the exact ground state $|\Psi_0\rangle$, rather on the uncorrelated HF one $|\Phi_0\rangle$ as in the usual CI calculations.

As was noted above, in our study the initially prepared non-stationary ionic state is created by suddenly removing an electron from a particular HF-orbital, i.e. by acting with the corresponding annihilation operator on the ground state of the neutral, $|\Phi_i\rangle = \hat a_i |\Psi_0\rangle$. Thus, only the 1h configurations contribute to the transition amplitude $\langle \Phi_i | I \rangle$, i.e. to the spectral intensity. Without correlation effects, the spectrum will consist of lines, one for every occupied orbital $\varphi_i$, with intensities equal to 1. If correlation effects are weak, the ionization spectrum will consist of \textit{main lines} which have large overlap with the 1h configurations. This is typical when ionizing the outer-valence shell of a system. In this case the molecular orbital picture is still valid. If the correlation effects are stronger, beside the main line \textit{satellite lines} will appear. The intensities of the satellite lines are weaker than those of the main lines since they correspond to cationic states that are dominated by 2h1p configurations and have only small or moderate overlap with the 1h configurations. Three types of satellites can be distinguished \cite{CedDoSchNie}: \textit{relaxation satellites} where at least one of the two holes in the 2h1p configuration is identical to the 1h orbital of the main line, \textit{correlation satellites} where both holes differ from the 1h orbital of the main line, and \textit{ground-state-correlation satellites} stemming from the correlation effects present in the ground state of the neutral. In the inner-valence, where the correlation effects are strong, the distinction between the main lines and the satellites cease to exist and the spectrum becomes a quasicontinuum of lines with small to moderate intensities. This phenomenon is known as breakdown of the molecular orbital picture \cite{CedDoSchNie}.

Depending on the structure of the ionic states involved, three basic mechanisms of charge migration have been identified \cite{Breidbach03}: the hole mixing case, the dominant satellite case, and the breakdown of the molecular orbital picture case. In what follows we will briefly describe them.

(\textit{i}) Hole mixing case. For simplicity, we will consider the two-hole mixing, i.e. the situation when two lines in the spectrum correspond to ionic states which are linear combinations of two 1h configurations $\hat a_j|\Psi_0\rangle$ and $\hat a_k|\Psi_0\rangle$. In this idealized case, if we create the initial hole in one of the orbitals, say $\varphi_j$, then the hole will oscillate between the two orbitals $\varphi_j$  and $\varphi_k$ with a frequency determined by the energy difference between the two ionic states. If the two orbitals are localized on two different sites of the system, the hole mixing mechanism will lead to an oscillation of the initially created positive charge between these two sites. The hole mixing mechanism was identified as the driving force of the ultrafast charge migration following outer- and inner-valence ionization in many different molecular systems (see, e.g. Refs. \cite{Hennig05,Kuleff05,Luennemann08,Luennemann_JCP08}).

(\textit{ii}) Dominant satellite case. Let us consider the situation when we have two ionic states, a main state and a satellite, both having overlap with the original 1h configuration. In CI language these states can be written as
\begin{eqnarray*}
|I_m\rangle=c_1\, \hat a_i|\Psi_0\rangle + c_2\, \hat a_a^\dagger \hat a_k \hat a_l |\Psi_0 \rangle,\\
|I_s\rangle=c_2\, \hat a_i|\Psi_0\rangle - c_1\, \hat a_a^\dagger \hat a_k \hat a_l |\Psi_0 \rangle,
\end{eqnarray*}
where the two coefficients $c_1$ and $c_2$ satisfy the equation $c_1^2+c_2^2=1$. In this idealized case, assuming that all involved orbitals $\varphi_i$, $\varphi_k$, $\varphi_l$, and $\varphi_a$ are different, i.e. the case of a correlation satellite, we will observe the following electron dynamics succeeding the ionization out of orbital $\varphi_i$. The hole initially localized on orbital $\varphi_i$ will migrate to the orbital $\varphi_k$ (or $\varphi_l$) accompanied by an excitation from orbital $\varphi_l$ (or $\varphi_k$) to the virtual orbital $\varphi_a$. Again the dynamics will be oscillatory with a period determined by the energy differences between the states $|I_m\rangle$ and $|I_s\rangle$. This type of mechanism was identified to be responsible for the ultrafast electron dynamics taking place after inner-valence ionization in several systems \cite{Kuleff05,Kuleff07,Holger2}. We note at this point that there is also a dominant \textit{relaxation} satellite mechanism which is not discussed so far and will be investigated for the first time in the present paper.

(\textit{iii}) Breakdown of the molecular orbital picture case. In the inner-valence region of the spectrum, where the quasicontinuum of lines appears, one can distinguish two general cases depending on whether the states are below or above the double ionization threshold of the system. Supposing that the quasicontinuum of states has a Lorentzian shape, in both cases the initially ionized orbital will ``lose'' its positive charge exponentially with time. If the states are below the double ionization threshold, the charge will be typically shared among many other orbitals and at the end of the process will be spread more-or-less uniformly over the whole cation. This situation was studied in Refs. \cite{Breidbach03,Holger2}. If the states are above the double ionization threshold, i.e. an electronic decay channel is open, this mechanism will describe the process of emission of a secondary electron. This situation was studied in Ref. \cite{tracing}.

In the present work we will concentrate on the dominant relaxation satellite mechanism of charge migration, and will discuss for the first time the manifestation of this mechanism after ionization of the outer-valence shell.

\section{Results}\label{sec_results}

We applied the methodology sketched in Sec. \ref{basics} to several molecules known to posses low-lying satellites in order to trace in real time and space the response of the electronic cloud of these systems to the sudden removal of an electron from their outer-valence shell. The systems studied were organic unsaturated nitroso compounds theoretically investigated by Wardermann and von Niessen \cite{Wardermann92}. The molecular geometries were optimized using DFT methodology (BP86/SV(P)). Throughout the remaining calculations standard DZ basis sets \cite{Dunning70} were used. All studied molecules show very similar behavior, thus in this paper we will present only one of them, namely the molecule 2-Nitroso[1,3]oxazolo[5,4-d][1,3]oxazole (see the sketch).
\begin{figure}[h]
\begin{center}
\includegraphics[width=2.5cm]{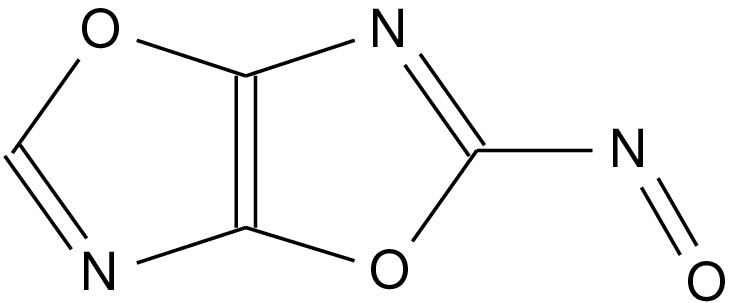}
\end{center}
\end{figure}

\begin{figure}[ht]
\begin{center}
 \includegraphics[width=6cm,angle=270]{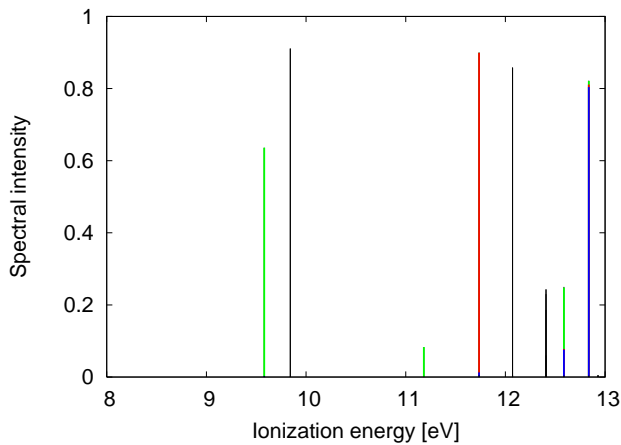}
\caption{Outer-valence part of the ionization spectrum of the molecule 2-Nitroso[1,3]oxazolo[5,4-d][1,3]oxazole calculated via the  non-Dyson ADC(3) Green's function method. Each vertical line shown is related to a final cationic state and is located at the corresponding ionization energy. The states belonging to the $a''$ irrep are plotted in black. All other colors are related to different 1h-configurations resulting from the removal of an electron out of a specific HF-orbital belonging to the $a'$ irrep.}\label{fig_spec}
\end{center}
\end{figure}

The molecule is planar and belongs to the $C_s$ symmetry point group having 29 occupied orbitals in the $a'$ irreducible representation (irrep) and 6 occupied orbitals in the irrep $a''$. The computed outer-valence ionization spectrum of the molecule is shown in Fig. \ref{fig_spec}. The calculations were performed via Green's functions (GF) based approach, namely the non-Dyson ADC(3) method \cite{Schirmer98}. The states belonging to the $a''$ irrep are plotted in black. All other colors are related to different 1h-configurations resulting from the removal of an electron out of a specific HF-orbital belonging to the $a'$ irrep. It is seen that the main line at 9.6~eV, the satellite at 11.2~eV and the larger contribution to the satellite at 12.6~eV come from the same 1h configuration, namely the $29a'^{-1}$ configuration which is given in green in the figure. The orbital $29a'$ is the highest occupied orbital (HOMO) of the molecule.

Let us take a closer look at the cationic states corresponding to these three lines in the spectrum, the main line and the two satellites. The first cationic state, corresponding to the line at 9.6~eV, is constructed mainly from the $29a'^{-1}$ 1h configuration (63\%) and from  the $29a'^{-1}6a''^{-1}7a''$ 2h1p configuration (25\%). The second state, corresponding to the satellite line at 11.2~eV, is also mainly constructed from the $29a'^{-1}$ and $29a'^{-1}6a''^{-1}7a''$ configurations but their weights are inverted compared to the first cationic state. The third state at 12.6~eV consists of 17\% $29a'^{-1}$, 10\% of $27a'^{-1}$, and 43\% of $29a'^{-1}6a''^{-1}7a''$. In all states the missing contributions are distributed over many other 1h and 2h1p configurations.

\begin{figure}[ht]
\begin{center}
\begin{tabular}{lll}
 a) & b) & c) \\
 & & \\
\includegraphics[angle=90,width=2.5cm]{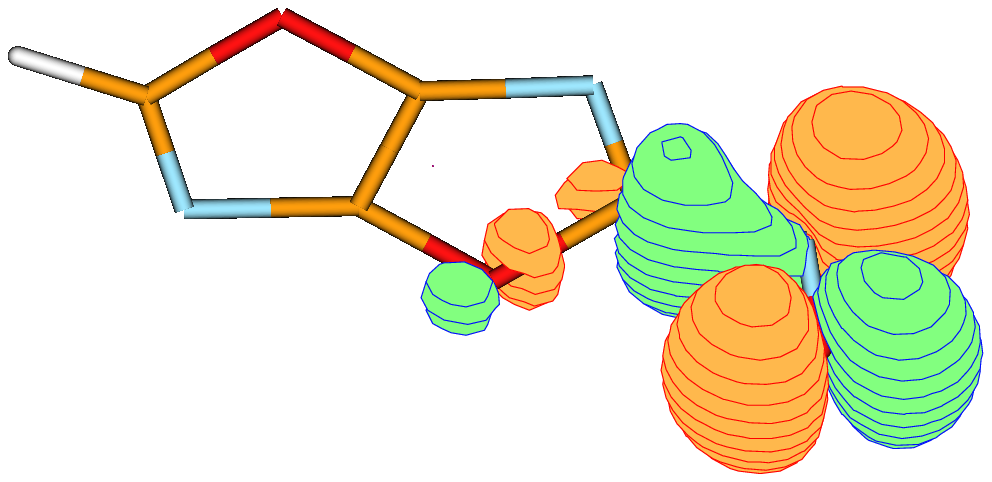}
\qquad
&
\includegraphics[angle=90,width=2.5cm]{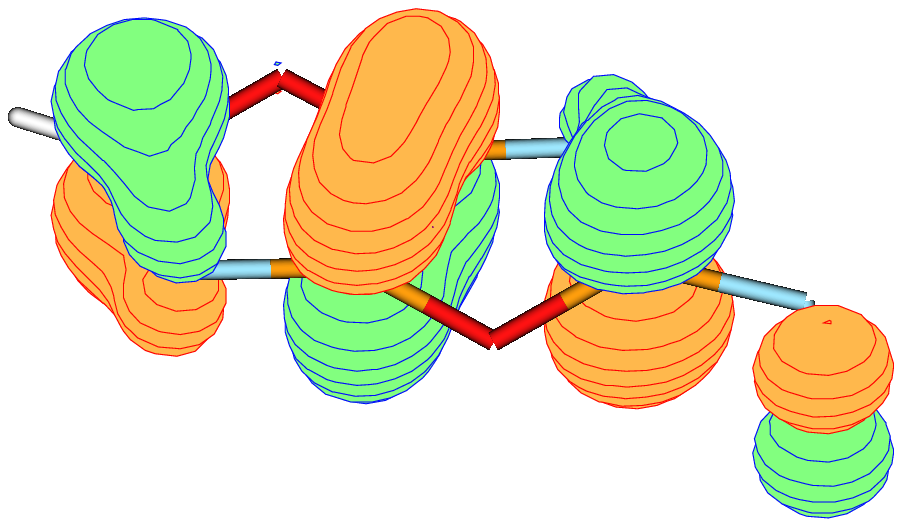}
\qquad
&
\includegraphics[angle=90,width=2.5cm]{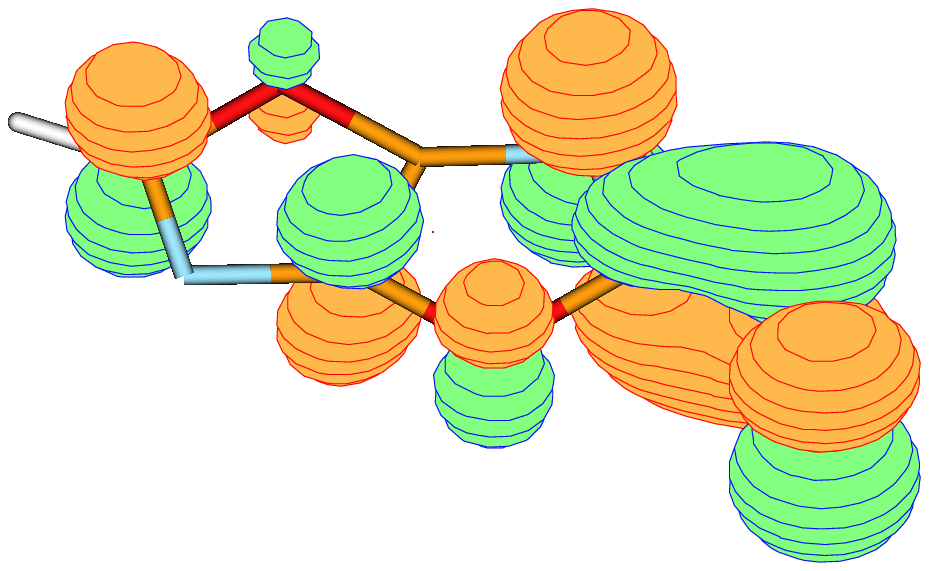}
\end{tabular}
\caption{Hartree-Fock orbitals of the molecule 2-Nitroso[1,3]oxazolo[5,4-d][1,3]oxazole: a) orbital $29a'$, or the HOMO; b) orbital $6a''$, or the HOMO-1; and c) orbital $7a''$, or the LUMO.}\label{fig_orbs}
\end{center}
\end{figure}

Obviously, we encounter the situation of a main line stemming from the ionization out of the HOMO and two relaxation satellites corresponding to an excitation of an electron from HOMO-1 to the lowest unoccupied molecular orbital (LUMO). The three involved HF-orbitals (the HOMO, orbital $29a'$, the HOMO-1, orbital $6a''$, the LUMO, orbital $7a''$) are displayed in Fig. \ref{fig_orbs}. It is seen that the HOMO is localized on the nitroso-site (N=O site) of the molecule, while the HOMO-1 and LUMO are delocalized $\pi$ orbitals.

\begin{figure}[ht]
\begin{center}
 \includegraphics[width=6cm,angle=270]{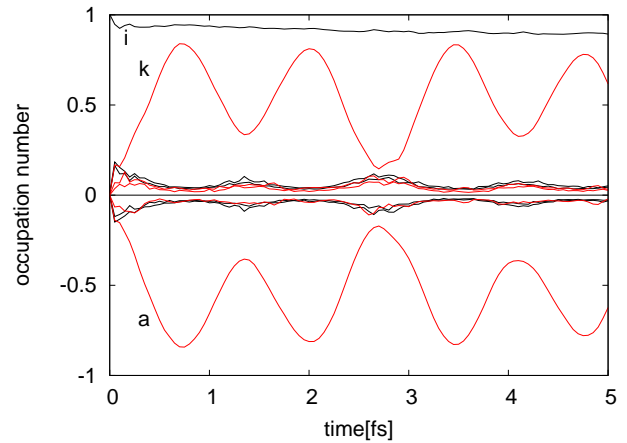}
\caption{Hole occupation numbers as a function of time for the first 5 fs after ionization of the HOMO (orbital 29a$'$) of the molecule 2-Nitroso[1,3]oxazolo[5,4-d][1,3]oxazole. The five most important occupation numbers of each irreducible representation are shown. At $t=0$ all occupation numbers are equal to 0 except of the one which relates to the initially ionized orbital. The negative hole occupations have to be interpreted as the corresponding fraction of an electron filling the respective virtual orbital.}\label{fig_occ}
\end{center}
\end{figure}

Let us now see what happens after a sudden removal of an electron out of the HOMO. In order to analyze the electron dynamics following the ionization it is illuminating to trace the time evolution of the hole occupation numbers $\tilde n_p(t)$. In Fig. \ref{fig_occ} we present the evolution of the hole occupation numbers during the first 5~fs after the ionization. The occupation numbers corresponding to orbitals belonging to irrep $a'$ are shown in black, while those corresponding to orbitals belonging to $a''$ in red. Only five of the occupation numbers from each irrep (those who contribute the most) are shown. At the beginning of the process all occupation numbers are equal to zero except one which corresponds to the initially ionized natural charge orbital $\tilde\varphi_i$, indicated by ``$i$'' in the figure. At that time the natural charge orbital $\tilde\varphi_i$ has 100\% overlap with the HF-orbital $29a'$. We see that apart from the very fast drop during the first 50 attoseconds, which is a universal response of a system upon sudden ionization (see Ref. \cite{Breidbach05}), the initially created hole charge stays in orbital $\tilde\varphi_i$ throughout the whole studied period. One should keep in mind that the natural charge orbitals also change with time, but in this case the orbital $\tilde\varphi_i$ varies very little and at any time point has more than 95\% overlap with the HF-HOMO. Thus, the positive charge practically stays where it is initially created.

However, another process takes place. We see that for approximately 0.7~fs more than 80\% of another hole appears on the natural charge orbital denoted by ``$k$'' in Fig. \ref{fig_occ}, and in the same time more than 80\% of an electron is promoted to the natural charge orbital denoted by ``$a$'' in the figure (note that ``negative'' hole occupation number describes a particle, i.e. an electron in a virtual orbital). The analysis shows that at all time points the natural charge orbital $\tilde\varphi_k$ overlaps more than 80\% with HF-orbital $6a''$, while the natural charge orbital $\tilde\varphi_k$ overlaps more than 60\% with HF-orbital $7a''$. Thus, the removal of an electron from the HF-HOMO leads to an ultrafast (less than a femtosecond) $\pi\rightarrow\pi^*$ excitation from orbital $6a''$ to orbital $7a''$. From Fig. \ref{fig_occ} we see also that this excitation process is cyclic, i.e. one observes alternating excitations and de-excitations with two beating periods one nearly twice as long as the other.

\begin{figure}[ht]
\begin{center}
\begin{tabular}{ccc}
\includegraphics[angle=90,width=2.5cm]{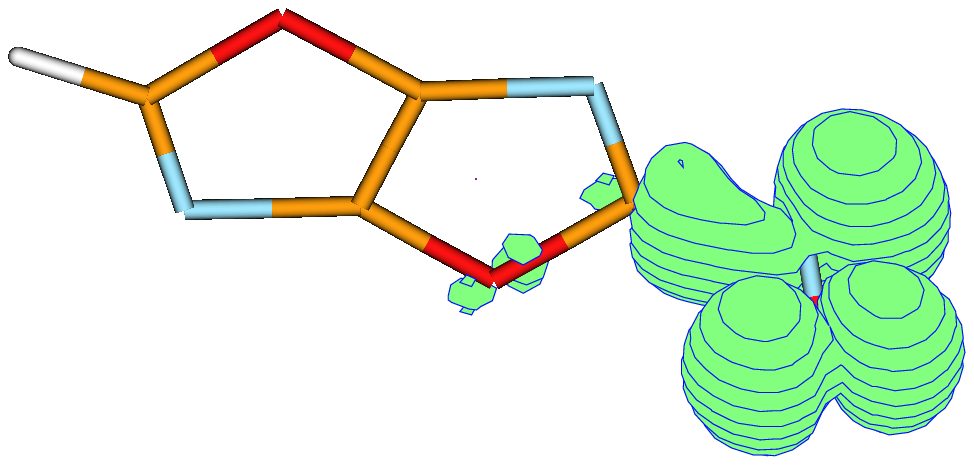} &
\includegraphics[angle=90,width=2.5cm]{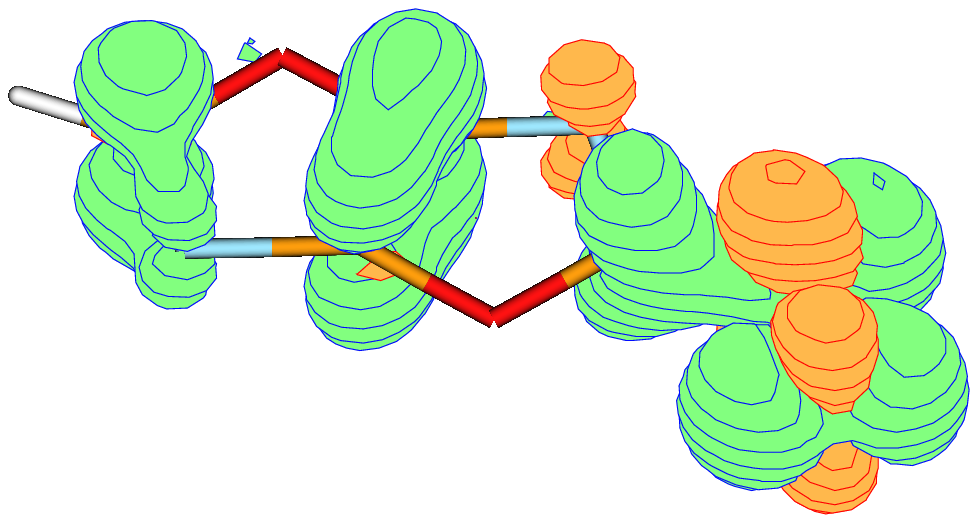} &
\includegraphics[angle=90,width=2.5cm]{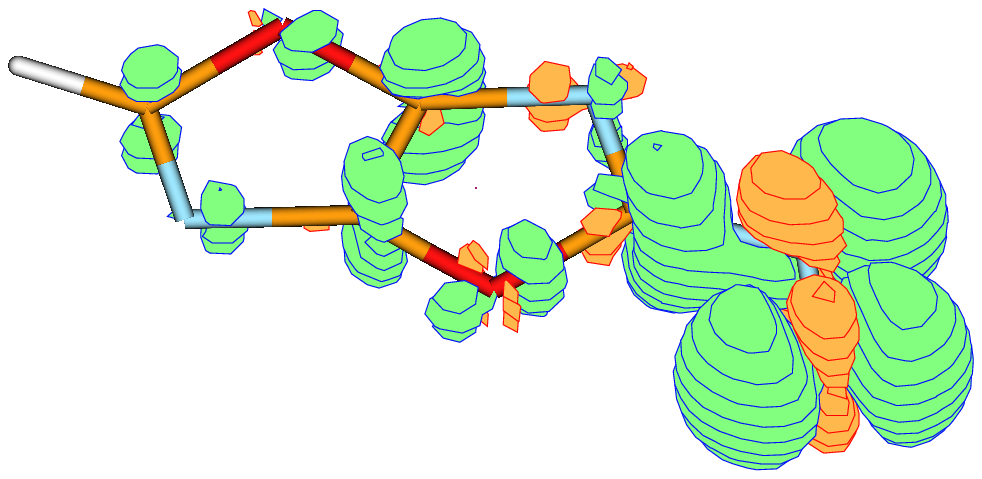} \\
0~fs & 0.7~fs & 1.4~fs
\end{tabular}
\begin{tabular}{cc}
\includegraphics[angle=90,width=2.5cm]{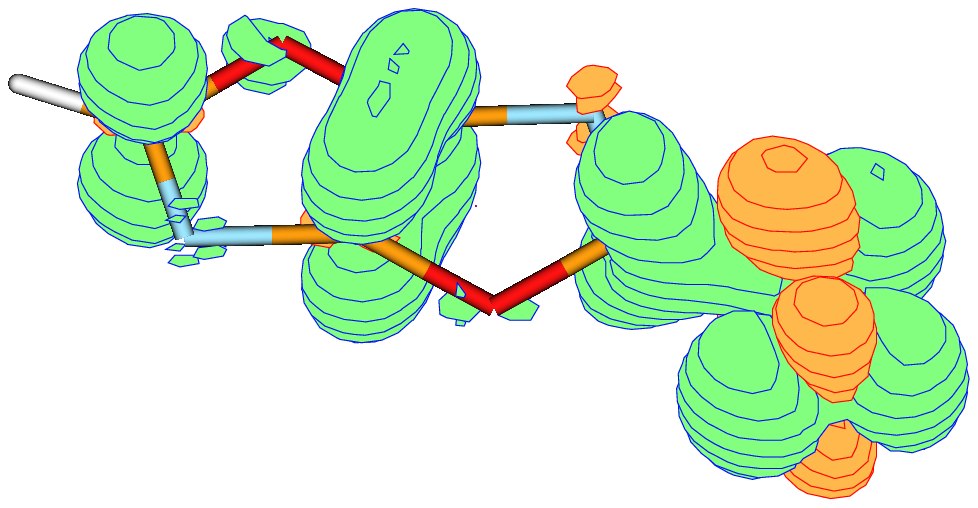} &
\includegraphics[angle=90,width=2.5cm]{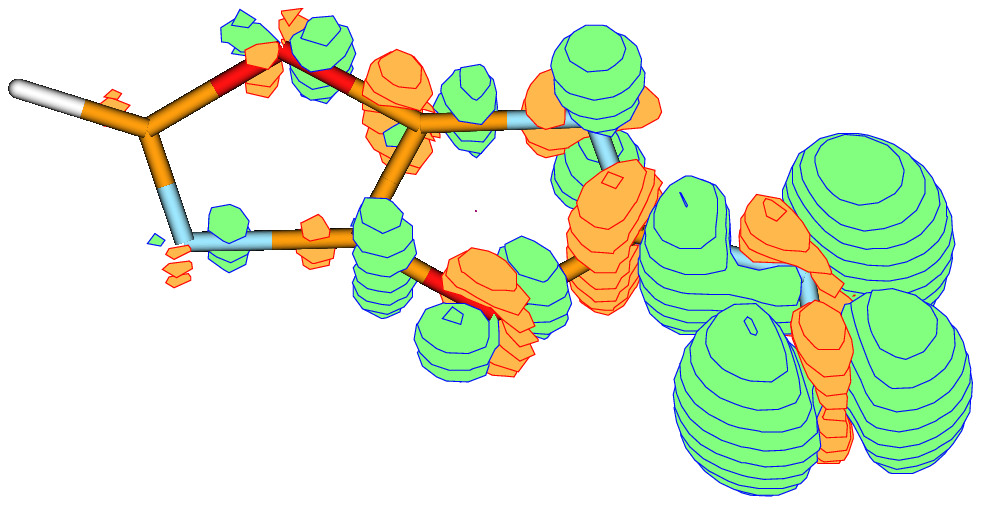} \\
2.1~fs & 2.8~fs
\end{tabular}
\end{center}
\caption{3D hole density $Q(\vec r,t)$ at times $t=$ 0, 0.7, 1.4, 2.1, and 2.8~fs after ionization of the HOMO of 2-Nitroso[1,3]oxazolo[5,4-d][1,3]oxazole. The ``negative'' hole density, or the electron density, is shown in orange.}\label{fig_dyn}
\end{figure}

Before analyzing in more detail the mechanism underlying this ultrafast electron dynamics, let us see how the process develops in space. For that purpose in Fig. \ref{fig_dyn} we display snapshots of the evolution of the hole density, Eq.~(\ref{eq_Q_n}), at different time points covering the time of a full oscillation cycle. The points chosen are $t=$ 0, 0.7, 1.4, 2.1, and 2.8~fs which correspond to the minima and maxima of the curves $k$ and $a$ in Fig. \ref{fig_occ}. We see that at $t=0$ the charge is localized on the nitroso side of the molecule (N=O group). As time proceeds a second hole starts to open at the left oxazole ring and an electron starts to appear at the nitroso group (shown in orange in Fig. \ref{fig_dyn}). One has to keep in mind that since the orbitals $6a''$ and $7a''$ are delocalized at the places of overlap we have a mutual cancellation between the electronic and hole densities. Thus, the process represents alternating ultrafast delocalization and localization of the charge. However, due to the large number of electronic states involved these alternations are not purely repetitive. From Fig. \ref{fig_dyn} it is clear that the charge tends to delocalize more and more as time proceeds.

\section{Analysis and discussion}\label{sec_discussion}

Here we would like to analyze in more detail the underlying mechanism of the ultrafast electron dynamics observed after sudden ionization of the HOMO of 2-Nitroso[1,3]oxazolo[5,4-d][1,3]oxazole. The dominant satellite mechanism studied in Ref. \cite{Breidbach03} and described briefly in Sec. \ref{CM_mechanisms} provides an understanding for the impact of a \textit{correlation} satellite on the electron dynamics following ionization. It leads to the migration of the initially created hole charge and a simultaneous excitation on top of it. However, as we saw in the previous section, in the case of dominant \textit{relaxation} satellite we observe only the excitation while the initially created charge remains stationary (see the red and black curves, denoted as $k$, $a$, and $i$, respectively in Fig. \ref{fig_occ}). This behavior can be understand with the help of a simple model.

Suppose we have only two cationic states -- a main state and a relaxation satellite. In CI language these states can be written as
\begin{eqnarray*}
 |I_m\rangle = c_1 |\Psi_i\rangle + c_2 |\Psi_{aik} \rangle,\\
 |I_s\rangle = c_2 |\Psi_i\rangle - c_1 |\Psi_{aik} \rangle,
\end{eqnarray*}
where $|\Psi_{i}\rangle \equiv \hat a_i|\Psi_0\rangle$ and $|\Psi_{aik}\rangle \equiv \hat a_a^\dagger \hat a_i \hat a_k|\Psi_0\rangle$. After some trivial algebra (see Ref. \cite{Breidbach03}) one arrives at the following analytical expression for the hole density matrix [Eq. (\ref{eq_N_t})]
\begin{eqnarray}\label{model_N}
&&\textbf N(t) = \nonumber \\
&&{\scriptsize\left(
\begin{array}{ccc}
 1 & 0 & 0 \\
 0 & -2(c_1c_2)^2[1-\cos(\omega t)] & -c_1c_2(c_1^2-c_2^2)[1-\cos(\omega t)] \\
 0 & -c_1c_2(c_1^2-c_2^2)[1-\cos(\omega t)] & 2(c_1c_2)^2[1-\cos(\omega t)]
\end{array}
\right),} \nonumber \\
\end{eqnarray}
where $\omega=(E_{I_s}-E_{I_m})/\hbar$.

The diagonalization of this matrix gives the time-dependent hole occupation numbers (using that $c_1^2+c_2^2=1$):
\begin{eqnarray}\label{model_occs}
 &&\tilde n_i(t)=1, \nonumber \\
 &&\tilde n_{k/a}(t)=\pm 2 c_1c_2\sin^2(\omega t/2).
\end{eqnarray}
We see that indeed the initially created hole is stationary and the occupations of the two other natural charge orbitals involved in the 2h1p configuration oscillate such that the total charge is equal to 1 at any time.

\begin{figure}[ht]
\begin{center}
\includegraphics[width=6cm,angle=270]{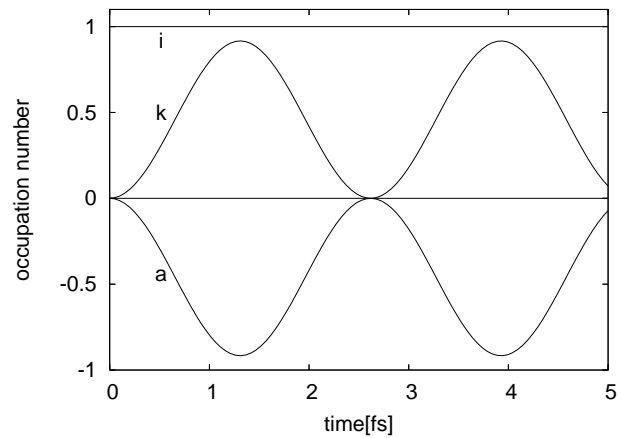}
\caption{Model time-dependent occupation numbers of the three natural charge orbitals ($i,k,a$) in the case of a dominant \textit{relaxation} satellite. The initially created hole ($i$) remains stationary, while an excitation from the occupied orbital ($k$) to the virtual orbital ($a$) is observed. The following parameters were used in the model: $c_1=\sqrt{0.7}$, $c_2=\sqrt{0.3}$, and $\omega=2.4$~fs$^{-1}$ [see Eq. (\ref{model_occs})].}\label{fig_occ_model}
\end{center}
\end{figure}

The time evolution of the three hole occupation numbers in this model are shown in Fig. \ref{fig_occ_model}. In order to compare with the full calculation presented in Fig. \ref{fig_occ}, the parameters used in the model are $c_1=\sqrt{0.7}$, $c_2=\sqrt{0.3}$, and $\omega=2.4$~fs$^{-1}$, which corresponds to the situation realized in the studied molecule taking only the main state at 9.6~eV and the first relaxation satellite at 11.2~eV. It is seen that this simple two-state model reproduces the overall behavior of the cyclic excitation and de-excitation with a period of about 2.6~fs. The more involved beating pattern observed in Fig. \ref{fig_occ} can be explained easily by the influence of the third state at 12.6~eV. 

It is straightforward to show that in the case of a main state and two relaxation satellites having the form
\begin{eqnarray*}
 |I_m\rangle = c^{(m)}_1 |\Psi_i\rangle + c^{(m)}_2 |\Psi_j \rangle+ c^{(m)}_3 |\Psi_{aik} \rangle,\\
 |I_{s1}\rangle = c^{(s1)}_1 |\Psi_i\rangle + c^{(s1)}_2 |\Psi_j \rangle+ c^{(s1)}_3 |\Psi_{aik} \rangle,\\
 |I_{s2}\rangle = c^{(s2)}_1 |\Psi_i\rangle + c^{(s2)}_2 |\Psi_j \rangle+ c^{(s2)}_3 |\Psi_{aik} \rangle,
\end{eqnarray*}
and supposing that $c^{(m)}_2=c^{(s1)}_2=0$, i.e. the 1h configuration $|\Psi_j \rangle$ does not contribute to the main state and one of the two satellites, the hole occupation numbers describing the excitation $k\to a$ will be proportional to $[\sin^2(\omega_1t/2)+\sin^2(\omega_2t/2)]$, where $\omega_2=(E_{I_{s1}}-E_{I_m})/\hbar$ and $\omega_1=(E_{I_{s2}}-E_{I_m})/\hbar$, while the orbital $j$ will get no occupancy, i.e. $\tilde n_j(t)=0$. This is exactly the case realized in the example studied in the present paper where the 1h configuration $|\Psi_j \rangle$ corresponds to $27a'^{-1}$ which contributes only to the second satellite state at 12.6~eV. It is seen that this second relaxation satellite will then just introduce an additional frequency $\omega_2=4.6$~fs$^{-1}$ corresponding to an oscillation period of about 1.4~fs and we will get the beating oscillation behavior observed in Fig. \ref{fig_occ}.

The following comment is in order. Although the relaxation satellite does not affect the dynamics of the initial hole charge itself (the initial hole remains stationary), depending on the spatial distribution of the involved orbitals the dominant relaxation satellite mechanism can lead to a charge migration via hole screening. If the electron is ejected from a localized orbital $i$ and the unoccupied orbital $a$ is localized in the same region of space, then the initially created hole will be screened by the excitation $k\to a$. Thus, in the idealized case when the orbitals $i$ and $a$ are localized on one site of the system, while the orbital $k$ is localized on a different site, the dominant relaxation satellite mechanism will lead to oscillations of the hole charge between these two moieties of the system. It has to be noted that the unoccupied orbitals are often delocalized and the realization of this simple mechanism of charge migration from one site to another is not very likely in such systems. This is also seen in the example presented in this work where the process represents rather an ultrafast delocalization and localization of the charge. However, in larger and/or strongly correlated systems where many orbitals can be mixed in the ionic states the hole-screening charge migration mechanism may be operative. We note that there do exist many systems with localized unoccupied orbitals, e.g. \textit{para}-Nitroaniline \cite{Lenz_nitroanil_CP79,Guerra_IJQC08}. Furthermore, in large systems even $\sigma^*$ orbitals are likely to be located on individual moieties. In all these kinds of systems one could expect ultrafast charge migration from one part of the system to another via the hole-screening excitation mechanism. 

Let us summarize. In the present work we studied the ultrafast electron dynamics following outer-valence ionization of a molecular system concentrating on the impact of low-lying relaxation satellites. For that purpose we traced the evolution of the electronic cloud after sudden removal of an electron from the HOMO of some organic unsaturated nitroso compounds known to possess low-lying satellites. Our results show that in all cases the initially created hole charge remains stationary but the system reacts by an ultrafast cyclic excitation-deexcitation process which leads to severe changes of the charge. In the presented example, the molecule 2-Nitroso[1,3]oxazolo[5,4-d][1,3]oxazole, the $\pi-\pi^*$ excitation following the removal of the HOMO electron takes place on a sub-femtosecond time scale, the period of the excitation-deexcitation alternations being about 1.4~fs. In real space the processes of excitation and de-excitation represent ultrafast delocalization and localization of the charge. In other systems with more localized $\pi^*$ or $\sigma^*$ orbitals one could anticipate by the same mechanism directed ultrafast charge migration from one site to another.

At the end we would like to note that this dynamical behavior of the electronic cloud can, of course, influence the nuclear dynamics which will come into play at later times since, at least within the Born-Oppenheimer approximation, the electronic motion governs the effective potential seen by the nuclei.

\begin{acknowledgments}
Financial support by the DFG is gratefully acknowledged.
\end{acknowledgments}

\end{document}